**Viral quasispecies profiles as the result of the interplay of competition and cooperation**

Juan Arbiza[1], Santiago Mirazo[1] and Hugo Fort [2†]

1. Instituto de Biología, Facultad de Ciencias. Iguá 4225. Universidad de la República. Montevideo, Uruguay.
2. Instituto de Física, Facultad de Ciencias. Iguá 4225. Universidad de la República. Montevideo, Uruguay.

† Corresponding author

e-mail addresses
    JA: jarbiza@fcien.edu.uy
    SM: smirazo@fcien.edu.uy
    HF: hugo@fisica.edu.uy.

**Abstract**

*Background.* Viral quasispecies can be regarded as a swarm of genetically related mutants. A common approach employed to describe viral quasispecies is by means of the quasispecies equation (QE). However, a main criticism of QE is its lack of frequency-dependent selection. This can be overcome by an alternative formulation for the evolutionary dynamics: the replicator-mutator equation (RME). In turn, a problem with the RME is how to quantify the interaction coefficients between viral variants. Here, this is addressed by adopting an ecological perspective and resorting to the *niche theory* of competing communities, which assumes that the utilization of resources primarily determines ecological segregation between competing individuals (the different viral variants that constitute the quasispecies). This provides a theoretical framework to estimate quantitatively the fitness landscape. *Results.* Using this novel combination of RME plus the ecological concept of niche overlapping for describing a quasispecies we explore the population distributions of viral variants that emerge, as well as the corresponding dynamics. We observe that the population distribution requires very long transients both to A) reach equilibrium and B) to show a clear dominating master sequence. Based on different independent and recent experimental evidence, we find that when some cooperation or facilitation between variants is included in appropriate doses we can solve both A) and B). We show that a useful quantity to calibrate the degree of cooperation is the Shannon entropy. *Conclusions.* In order to get a typical quasispecies profile, at least within the considered mathematical approach, it seems that pure competition is not enough. Some dose of cooperation among viral variants is needed. This has several biological implications that might contribute to shed light on the mechanisms operating in quasispecies dynamics and to understand the quasispecies as a whole entity.

**Background**

The concept of quasispecies [1] refers to the equilibrium spectrum of closely related mutants, dominated by a master sequence, generated by a mutation-selection process. It has become an adequate descriptor of RNA viruses at the population level [2,3] and provides natural links between population biology and virology.

Competitive exclusion and displacement when two viral populations with nearly equal starting fitness compete has been observed [4]. Thus, a challenging puzzle is how so many (sometimes very similar) variants can coexist in nature. In the case of ordinary ecosystems niche differentiation is obviously an important aspect to promote biodiversity [5]. However, the often striking similarity in coexisting variants suggests that other mechanisms must be involved.

An infected individual can harbor several genetically related variants of the same species (assuming that this concept is valid for RNA viruses), and thus the host can be regarded as an ecosystem in which distinct viruses interact.

These interactions are in general quite complex and may be direct or indirect -for example, when the host's immune system responds against one particular viral variant, affecting the whole fitness landscape. There is a process called *complementation,* in which one virus provides *in trans* a useful product that cannot be made by another variant. On the other hand, it was proposed that *in trans* acting defective gene products, expressed by mutagenized viruses can interfere with replication and prevent replication of pathogenic viruses [6]. Thus, under such circumstances viral phenotype may not necessarily reflects viral genotype.

Obviously, if the viruses provide each other with useful resources the interaction is of mutual benefit. Indeed this is what it was found in the evolution of competitive interactions among RNA phage viruses [7]. When two mutants co-infect a cell the common resource pool allows the viruses to use each other's protein products. The co-infection rescues the mutants, allowing them to reproduce when they would be otherwise unable to do so [8,9].

Furthermore, it was recently found that for polio viruses the diversity of the quasispecies population, rather than selection of individual adaptative mutations, may determine pathogenesis through cooperative interactions [10,11].

One of the common approaches employed to describe viral quasispecies is by means of the *quasispecies equation* [12] (QE) which is limited by the fact that it lacks frequency-dependent selection. In other words, the fitness of a particular phenotype is set to a constant value independently of the other competing `players´. This can be fixed by using the *replicator-mutator equation* [13] (RME), which represents a general formulation for the evolutionary dynamics.

A far from trivial problem with the RME is how to get the interaction coefficients among variants. Here we adopt an ecological perspective and resort to the *niche theory* [14] of competing communities, which assumes that exploitation (or utilization) of `resources´ primarily determines ecological segregation. Thus, we regard the swarm of mutants that constitute a quasispecies as a set of interacting RNA molecules distributed along a hypothetical *niche axis* [15]. This allows, in a simple way, to estimate the size of the interaction coefficients of the RME. Basically, the degree of overlap between variants is what ultimately determines the intensity of these coefficients (see Method section).

The other novel ingredient, besides the use of niche theory, is to consider the effect of some measure of cooperation or facilitation among viral variants. We start by showing that in general niche theory does not lead by itself to a `typical´ quasispecies profile (see Results section). Then we show that this problem can be solved if some dose of cooperative interactions between variants is taken into account. Therefore, in order to get a `typical´ quasispecies profile both competition and cooperation or facilitation between viral variants is needed.

**Results**

In the case of pure competition, i.e. all interaction coefficients $\alpha_{ij}$ negative (see Method section), we obtain distributions like the ones shown in Figure 1, which represent in each row the evolution for a different value of $\sigma$: $\sigma = 0.2$, $\sigma = 0.15$ and $\sigma = 0.1$ in rows 1, 2 and 3, respectively. The left panels are histograms of the fractions $x_i$ of phenotype $i$ vs. their corresponding positions on the niche axis $\mu_i$ for $N_g$ =500 (see Method section for a detailed description of parameters). In the three cases, a pattern of lumps -2, 3 and 4 respectively- separated by gaps along the niche axis emerges. This multi-lump population profile is in complete agreement with the one recently found by Scheffer and van Nes [16] for L-V competition models, dubbed by them as *self-organized similarity* (SOS). This is not surprising since the L-V competition model is mathematically equivalent to a replicator dynamics. In fact, up to now, the main difference between both treatments is in the way mutations are implemented, but this seems not to affect qualitatively the results. Furthermore, the observed inverse dependence in the number of lumps with σ was recently explained [17]. The right panels in Figure 1 show the temporal evolution of each fraction during the $N_g$ =500 generations.

Notice that this multi-lump pattern resembles several co-infecting quasispecies (one per lump) rather than a population profile of a single quasispecies. Moreover, after 500 generations, a) there are many phenotypes that survived, and b) the viral ecosystem has not reached equilibrium yet (the population fractions of several variants are still considerably changing with time). So the pure competitive niche model does not seem to work properly.

Let us now explore what happens when some degree of cooperation or facilitation between variants is included in this model by turning a random fraction $f_C$ of the interactions from competitive to cooperative (see Table 1 in Method section). Figure 2 is completely analogous to Figure 1 but when 1% ($f_C$ = 0.01) of all $\alpha_{ij}$ are positive, illustrating the dramatic changes introduced by taking into account even a very small amount of cooperation. First of all, notice that the multi-lump pattern disappeared and only some variants remain. This is particularly clear for $\sigma = 0.2$ and $\sigma = 0.15$ with a master sequence that represents around 50 and 40 % of the population, respectively. In addition, the population profile does not change very much after a few hundred generations, in agreement with the concept of a quasispecies. In contrast, if $\sigma$ is too small (Figure 2 row 3) the equilibrium can not be easily attained, and therefore these results offer a criterion to estimate a lower limit for this parameter.

It is also interesting to evaluate the dependency on $f_C$. Consider, for example, increasing it one order of magnitude, from 0.01 to 0.1. In this case there are no qualitative changes. However, if we increase the degree of cooperation even further to $f_C = 0.5$ (50 % of cooperative interactions) and the other two parameters are set to $\sigma = 0.15$ and $P_m$=0.1, the whole picture changes again. Figure 3 clearly shows that after a few hundred generations, much more variants survive, the master sequence is below 14 % of the total population and still the populations are changing drastically. That is, the profiles for large values of $f_C$ become widely distributed, with a master sequence that seems under-represented. In order to further address this question we have performed, for different values of $f_C$ (and always for $\sigma = 0.15$), 1000 simulations corresponding to 1000 distinct realizations. Each realization corresponds to a different random distribution of variant populations, and consequently different interaction matrices $\alpha_{ij}$. Moreover, different growth rates $r_i$ were assigned. We have computed two quantities by averaging over these 1000 realizations: First, $\bar{p}_{1/2}$, the number of variants, among the $N = 100$ considered, which after $N_g$ =500 generations represent at least 1/2 of the master sequence (of course 1/2 is an arbitrary threshold). Second, another measure of variability given by the normalized Shannon entropy (S) calculated as $S_n = - [\Sigma_i x_i \ln(x_i)]/\ln(N)$. The possible values of $S_n$ range from zero (when there is just one variant, the master) to one (when all variants differ from one another and are equally represented). For normalized Shannon entropy applied to viral quasispecies see, for instance, reference [18].

We found that these average percentages $\bar{p}_{1/2}$ vary from 1.6 % for $f_C = 0.01$ to 5.3 % for $f_C = 0.5$ (see Table 2). Since we don't have experimental data of quasispecies profiles with such detail, it is difficult to determine a critical value for $f_C$, $f_C^*$. However, the scant experimental data seem to favor small values of $\bar{p}_{1/2}$.

In the case of $S_n$ we have found that the average $\bar{S}_n$ varies from 0.474 for $f_C = 0.01$ to 0.471 for $f_C = 0.1$ and to 0.592 for $f_C = 0.5$, while in the case without cooperation, $\bar{S}_n = 0.708$. Therefore, the case in which $f_C = 0.5$ appears to be closer to the purely competitive situation, indicating again too much variability. So these results also favor smaller values of $f_C$.

**Discussion**

In this study we have modeled quasispecies by the RME plus the ecological concept of niche overlapping to quantify the intensity of interactions within a quasispecies. In other words, the swarm of viral variants that constitute a quasispecies is regarded as a set of interacting viral variants distributed along a niche axis.

The population distributions that emerge when only competition between variants is taken into account are different from what one would expect for a quasispecies profile: they are very unstable and resemble more to several coexisting quasispecies than to just a single one. Both problems can be solved if some dose of cooperation or facilitation between variants is included. It is worth stressing that this is just one possible solution to get a population profile that resembles that of a quasispecies from a general evolution model. The existence of cooperation among viruses is supported by recent experimental data, both in the case of co-infection [8,9] and among the variants that constitute a quasispecies [10]. In fact, the composition of a quasispecies mutant spectrum may indeed determine the viral population behavior through complementation or interference, i.e., positive and negative interactions, respectively [10]. It was recently demonstrated that interfering potential of replication-competent mutants may eventually modulate viral infection and population behavior of highly variable RNA viruses [19].

In addition, we have found two, at first sight, surprising results. First, the cooperation among variants seems to undermine, rather than enhance, biodiversity: much more variants within a quasispecies survive in the case of pure competition between them than when some facilitation occurs. Second, the effect of changing a small percentage of interactions from competitive to cooperative is more drastic than changing a larger fraction. This becomes evident from the comparison between results for $f_C = 0.01$ with those for $f_C = 0.5$ cooperative interactions. These two outcomes can be understood, from a mathematical point of view, as follows. When a small number of positive interactions, completely at random, are incorporated, the fitness of the few ´lucky variants`, which receive help from others or become involved in cooperative cycles, soars. This has a strong destabilizing influence and, in consequence, many variants or RNA molecules disappear. So the whole effect is that diversity decreases. As the percentage of cooperative interactions is increased, the balance is gradually restored and so is diversity. This is also consistent with $\bar{p}_{1/2}$ and $S_n$ analysis (see Table 2); the higher the value of the proportion of cooperative interactions $f_C$, the higher the value of these two parameters. These data offer a guide to estimate the order of the percentage of cooperation that is needed to reproduce a quasispecies profile. In fact, Figure 2 fits better with the idea one has of a quasispecies than Figure 3.

One might speculate that the profile of randomly assigned growth rates $r_i$ can have an important effect in the above results. However, we checked that with a uniform profile i.e. $r_i = r = 1$ for all variants results are qualitative similar. Thus in general the major role in determining the fitness of each variant comes mostly from the interaction with other variants.

The above discussed results have several direct biological implications that may contribute to shed light over the mechanisms operating in population dynamics, particularly in quasispecies mutant spectra behavior. As it is known, in an infected organism during normal infection, multiple viral genomes compete for resource needed to complete life cycle. Due to the high mutation rate of RNA viruses [20], defective mutants or genome defectors coding for non-functional proteins able to use these resources arise, likely interfering with efficient replication of viable viruses. In fact, what lethal defection model proposes is that when a slight increase in the level of mutant defectors occurs (e.g., by mutagenic activity), viral extinction may be achieved [21].

**Conclusions**

In this work we introduce two main novelties to mathematically describe the dynamics of RNA viral populations. First, we resort to niche theory in order to quantitatively estimate the interactions between viral variants constituting a quasispecies. Second, we include cooperative interactions between these viral variants and analyze their effects from an evolutionary point of view.

The model we propose is constituted then by two main parameters: the overlapping parameter $\sigma$, which controls the intensity of the interactions, and the percentage of cooperative or positive interactions $f_C$, which determines the sign balance of these interactions (see Method section).

Concerning potential applications, this model could be eventually applied to better understand complexity of viral population behavior in the course of chronic infections (e.g. those caused by Hepatitis C Virus and Human Immunodeficiency Virus).

Finally, it is important to stress that in the RME model diversity by itself cannot promote facilitation or cooperation since the interaction matrix do not evolve. In fact, as in the canonical RME model there is always just competition. Therefore we explicitly introduce some dose of cooperation measured by $f_C$. This is a nuance to ref. [10] in which the whole quasispecies diversity emerges as relevant in evolutionary terms. An evolving interaction matrix from which cooperation might emerge would imply a different and quite more complex model. Indeed this is a very interesting aspect concerning viral evolution, which is beyond the scope of this work but worth analyzing in the future.

## Method

### Model

We approach the quasispecies as an ecosystem, what implies that the fitness of each viral phenotype or virus variant results from the combination of its intrinsic properties (collected in a maximum growth rate parameter) and those resulting from its interactions (interaction coefficients, negative in the case of competition and positive for cooperation) with the other variants. Thus, in a very simplified way, each variant is represented by a single quantity which can be thought as an aggregate of several relevant properties like the affinity of the virus variant to bind to a cell receptor, its resistance to interferon, etc. In other words we are considering a hypothetical continuous one-dimensional *niche axis*. One may think in the case of ordinary ecosystems, involving plants or animals, that the `position´ of a species on this niche axis is related to its body size.

### Replicator Dynamics plus Mutations

We denote by $x_i$ the relative abundance or frequency of phenotype *i* and then we have $\sum_{i=0}^{N} x_i = 1$. The fitness of phenotype *i* is denoted by $f_i$, a non-negative real number corresponding to the rate at which the phenotype *i* replicates. The average fitness of the population is given by $\bar{f} = \sum_{i=0}^{N} x_i f_i$. During replication of a genome, mistakes can happen. The probability of replication of a variant *i* results in variant *j* is given by $q_{ij}$ and then these mutation coefficients obey $\sum_{j=1}^{N} q_{ij} = 1$. Since $q_{ii}$ is the probability that the variant doesn't change by mutation, the probability that it does is given by 1- $q_{ii}$. A common equation used to model replication of viruses is the *quasispecies equation*:

$$\frac{dx_i}{dt} = \sum_{j=1}^{N} x_j f_j q_{ji} - x_i \bar{f} \qquad i = 1,...,N . \qquad (1)$$

A problem with this equation is that the fitness of a particular type is set to a constant value. We want to allow the fitness landscape to incorporate the distribution of the population variants rather than setting the fitness $f_i$ of a particular variant constant. Alternatively, one could treat the swarm of virus variants as an ecosystem and use the Lotka-Volterra competition (L-V) equations:

$$\frac{dx_i}{dt} = r_i x_i (1 - \sum_{j=1}^{N} \alpha_{ij} x_j) \qquad i = 1,...,N , \qquad (2)$$

where $r_i$ is the maximum per capita growth rate and the coefficients $\alpha_{ij}$ represent the interaction of variant *j* over variant *i*. These *N* L-V equations are in turn mathematically equivalent to a set of *N+1* *replicator* equations [22] given by:

$$\frac{dx_i}{dt} = x_i(f_i - \bar{f}) \qquad i = 1,...N+1$$

$$f_i = r_i \sum_{j=1}^{N+1} \alpha_{ij} x_j ,$$

(3)

where $\bar{f}$ is the average value of the fitness $\bar{f} = \sum_{j=1}^{N+1} f_j x_j$ , provided $\alpha_{N+1 j} = 0$ for $j = 1,...,N+1$, and $\alpha_{iN} = r_i$ for $i = 1,...,N$.

The replicator equation succeeds in incorporating to the fitness landscape the population distribution of variants but has the problem that, since it does not incorporate mutations, it is non-innovative. Hence we will resort to a hybrid model, a kind of replicator with mutations or *replicator-mutator* equations, given by

$$\frac{dx_i}{dt} = r_i \sum_{j=1}^{N+1} \sum_{k=1}^{N+1} \alpha_{jk} (q_{ji} - x_i) x_j x_k \qquad i = 1,...,N+1$$

$$\alpha_{N+1 j} = 0 \qquad j = 1,...,N+1$$

$$\alpha_{iN} = 1 \qquad i = 1,...,N,$$

(4)

to describe N virus variants evolving by selection and mutations. On the one hand, if there were no mutations, *i.e.* $q_{ij} = 1$ if *i=j* and 0 otherwise, equation 4 would reduce to the well known replicator equations. On the other hand, by replacing $\sum_{j=1}^{N+1} \alpha_{ij} x_j$ by a fixed $f_i$, we would recover the quasispecies equations. Therefore, this hybrid model generalizes the quasispecies and replicator equations joining the advantages of both of them: selection and innovation. Moreover, previously reported RME describing frequency dependant selection and mutation, has been used in population genetics [23] and language evolution [24].

**Niche overlap, interaction coefficients and mutation probabilities**

To compute the coefficients $\alpha_{ij}$, adopting an ecological point of view, let's consider that the interaction between species (virus variants in our case) depends on the *niche overlap*. That is, on the `proximity´ along the niche axis: the closer the stronger. A given variant *i* is represented by a certain probability distribution position on the niche axis ($\xi$): $P(\xi)$ around its average position $\mu_i$.

A reasonable assumption for the shape of $P(\xi)$ is the normal distribution $P(\xi) = \frac{1}{\sqrt{2\pi}} e^{-(\xi-\mu)^2/(2\sigma^2)}$. To avoid border effects, the niche axis is defined circular, i.e. we impose periodic boundary conditions, so that each variant has equal numbers of competitors on both sides. The intensity of the interaction between variants $i$ and variants $j$ is related to niche overlap, and thus to the probability $P$ that individuals of the two variants are at the same position on the niche axis, which is the product of both probabilities. For two variants with probability distributions centered at $\mu_1 = 0.4$ and $\mu_2 = 0.6$ and both with $\sigma = 0.1$, a schematic picture is given (Figure 4)

Hence the absolute values of the interaction coefficients $|\alpha_{ij}|$ can be expressed as the ratio of the probability of matching an individual of variant $j$ and the probability of matching one of the same variant, that is [15]:

$$|\alpha_{ij}| \equiv \frac{\int_{-\infty}^{+\infty} P_i(\xi) P_j(\xi) d\xi}{\int_{-\infty}^{+\infty} P_i^2(\xi) d\xi} = e^{-\frac{(\mu_i - \mu_j)^2}{4\sigma^2}} \qquad i=1,\ldots N-1;\ j=1,\ldots N-1, \qquad (5)$$

(hence $\alpha_{ii} = 1$). The sign of the $\alpha_{ij}$ is negative or positive depending if the interaction is competitive or cooperative, respectively.

We assume that the mutation probabilities $q_{ij}$ also depend on the proximity along the niche axis, and then they are proportional to the absolute value of the $\alpha_{ij}$ for the case of $i$ different from $j$. The proportionality factor $m$ is chosen in such a way that the probability of an error in each variant replication $P_m \equiv 1 - \langle q_{ii} \rangle$ where the brackets $\langle .. \rangle$ denoting the average over the population, is of the order of 10 %. In any case, we checked that the model is quite robust against variation of this parameter. Figure 5 shows the evolution of the population profile for $\sigma = 0.15$ after 250 and 500 generations. From panels (A) to (C) it becomes clear that the initial phenotype is disappearing (indeed after 1000 generations it becomes 0 for all practical purposes). Moreover, we checked that the emergence of this pattern is independent from the chosen initial configuration, e.g. the same occurs if one takes an arbitrary distribution of random `seeds´ consisting in many different viral phenotypes distributed along the niche axis. In this case the diagonal probabilities $q_{ii}$ are computed using the normalization condition $\sum_{j=1}^{N} q_{ij} = 1$.

We proceed by partitioning the segment [0-1] representing the niche axis into $N$=100 sub-segments each corresponding to a given virus variant, i.e. the viral phenotypes are binned into $N$=100 categories. We start with these variants uniformly distributed at positions $\mu_i$ on the niche axis, each with the same niche width given by the standard deviation $\sigma$ (typically $\sigma$ =0.15). Then we let the system evolve over $N_g$ generations according to eq. 4.

Initially we consider the case of pure competition, i.e. all the $\alpha_{ij}$ are negative. Next we allow some dose of cooperation or facilitation between species. We implement this in the simplest way: a fraction $f_C$ of the interactions, chosen at random, turns from competitive to cooperative i.e. if some of the $\alpha_{ij}$ change of sign and become positive. As in the case of pure competition the whole interaction matrix remains fixed during evolution. Thus in particular we want to stress that the interaction between two given variants doesn't switch between being cooperative and defective during the evolution of the quasispecies.

See Table 1 for brief description of parameters involved in the model.

**Authors' Contributions**

HF and JA conceived and designed the model and experiments. HF, JA and SM analyzed the data and wrote the paper. All authors read and approved the final manuscript.

**Acknowledgments.**

We would like to thanks Esteban Domingo for useful discussion and critical reading of this manuscript. We also would like to thanks administrative and financial support from PEDECIBA and Agencia Nacional de Investigación e Innovación.

**References**

1. Eigen M and Schuster P: The Hypercycle: a principle of natural self-organization. Springer-Verlag 1979. New York.

2. Domingo E, Escarmis C, Sevilla N, Moya A, Elena SF, Quer J, Novella IS and Holland J: Basic concepts in RNA virus evolution. *FASEB Jour* 1996. 10; 859-864.

3. Eigen M and Biebricher CK: Sequence space and quasispecies distribution. In RNA Genetics (Domingo E, Holland J, and Ahiquist P. eds) 1998. Vol. 3, pp.211-245. CRC Press.Inc., Boca Raton, Florida.

4. Clarke D, Duarte E, Elena S, Moya A, Domingo E and Holland J: The red queen reigns in the kingdom of RNA viruses. *PNAS* 1994. 91; 4821-4824.

5. May RM: Stability and complexity in model ecosystems, Princeton; Princeton University Press; 1974.

6. González-López C, Arias A, Pariente N, Gomez-Mariano G and Domingo E: Preextinction viral RNA can interfere with infectivity. *J. Virol.* 2004. 78 (7); 3319-3324.

7. Turner PE and Chao L: Sex and the evolution of intrahost competition in RNA virus $\phi$6. *Genetics 1998.* 150;523-532.

8. Turner PE and Chao L. Escape from prisoner's dilemma in RNA phage $\phi$6. *Am. Nat.* 2003. 161 (3); 497–505.

9. Turner PE: Prisoner's dilemma in an RNA virus. *Nature* 1999. 398; 441–443.

10. Vignuzzi M, Stone JK, Arnold J, Cameron C and Andino R: Quasispecies diversity determines pathogenesis through cooperative interactions in a viral population. *Nature* 2006. 439; 344-348.

11. Pfeiffer J, and Kierkegaard K : Increased fidelity reduces poliovirus fitness and virulence under selective pressure in mice. *PLoS Pathog.* 2005. 1(2):e11; 102-110.

12. Nowak MA: Evolutionary Dynamics: Exploring the Equations of Life. Cambridge, USA: Harvard University Press; 2006.

13. Page KM and Nowak MA: Unifying evolutionary dynamics. *J. Theor. Biol.* 2002. 219; 93-98.

14. Levins R: Evolution in Changing Environments. Princeton: Princeton University Press;1968.

15. MacArthur RH and Levins R: The limiting similarity, converge, and divergence of coexisting species. *Am. Nat.*1967. 101; 377–385.

16. Scheffer M and van Nes E. Self-organized similarity, the evolutionary emergence of groups of similar species. *PNAS. USA.*2006. 103; 6230-6235.

17. Fort H, Scheffer M and van Nes E: The paradox of the clumps mathematically explained. *Theor. Ecol.* 2009.2 (3); 171-176.

18. Sierra S, Dávila M, Lowenstein P and Domingo E: Response of Foot-and-Mouth disease virus to increased mutagenesis: influence of viral load and fitness in loss of infectivity. *J Virol.* 2000. 74(18): 8316–8323.

19. Perales C, Mateo R, Mateu M and Domingo E: Insights into RNA virus mutant spectrum and lethal mutagenesis events: replicative interference and complementation by multiple point mutants. *J.Mol.Biol.* 2007. 369; 985-1000.

20 . Drake JW and Holland JJ: Mutation rates among RNA viruses. *PNAS* 1999. 96(24); 13910-13913.

21. Grande-Pérez A, Lázaro E, Lowenstein P, Domingo E and Manrubia S: Suppression of viral infectivity through lethal defection. *PNAS* 2005. 102; 4448-4452.

22. Hofbauer M and Sigmund K: Evolutionary Games and Population Dynamics. Cambridge, UK: Cambridge University Press; 1998.

23. Hadeler KP: Stable polymorphisms in a selection model with mutation. *SIAM J. Appl. Math.* 1981. 41; 1–7.

24. Nowak, MA, Komarova N L and Niyogi P: Evolution of universal grammar. *Science.* 2001. 291;114–118.

**Figures**

**Figure 1 - Population fractions for the case of pure competition.**

Figure1. Relative abundance of each variant after 500 generations for, respectively, σ = 0.2, σ = 0.15 and σ = 0.1. Corresponding temporal evolution of the fractions of each variant is shown (left panels).

**Figure 2 -Population fractions for the case $f_C$ = 0.01.**

Figure 2. Analogous to Figure1 for the case of $f_C$ = 0.01. Relative abundance of each variant after 500 generations for $\sigma$ = 0.2, $\sigma$ = 0.15 and $\sigma$ = 0.1. Corresponding temporal evolution of the fractions of each variant is also shown.

**Figure 3 - Population fractions when half of the interactions are cooperative.**

Figure 3. Relative abundance of each variant after 500 generations (Left) and temporal evolution of the fractions of each variant (Right).

**Figure 4 - RNA molecules niche position ($\xi$) and overlapping.**

Figure 4. Scheme showing two normal probability distributions centered at $\mu_1$ = 0.4 and $\mu_2$ = 0.6 of two RNA molecules, both with $\sigma$ = 0.1. The region in black corresponds to the overlap between the two normal distributions.

**Figure 5 - The independence from the initial state.**

Figure 5. Panel (A) shows the evolution of the population profile for $\sigma$ = 0.15 starting from only one viral variant at an arbitrary position $\mu_j$ on the niche axis (that is, $x_i$=0 for all i ≠ j and $x_j$ = 1). Panel (B) and (C) corresponds to the population profile after 250 and 500 generations, respectively. Emergence of three different lumps becomes clearer after 500 generations.

**Tables**

Table 1. Parameters involved in the model

| PARAMETER | Controls | Typical values |
|---|---|---|
| Standard dev. of pop. distributions $\sigma$ | Niche overlap | 0.1-0.2 |
| Mutation proportionality factor $m$ | Phenotype Mutation Rate $P_m$ per generation | 0.0035-0.002 so that $P_m$ =0.1 |
| Maximum per capita growth rate $r_i$ | Intrinsic growth rate | Random number in [0,1] |
| Fraction of positive interactions $f_C$ | Degree of cooperation | 0.01 |

Implicancies of each parameter included in our model is listed; the quantities they control and their typical values.

Table 2. Role of $f_C$ in the model.

| $f_C$ | $\overline{p}_{1/2}$ | $S_n$ |
|---|---|---|
| 0.01 | 1.6 | 0.474 |
| 0.1 | 3.0 | 0.471 |
| 0.25 | 3.7 | 0.505 |
| 0.5 | 5.3 | 0.592 |

$\overline{p}_{1/2}$ -the average % of viral variants whose population represent at least ½ of the master sequence- for different $f_C$ and normalized Shannon entropy $S_n$ is shown.

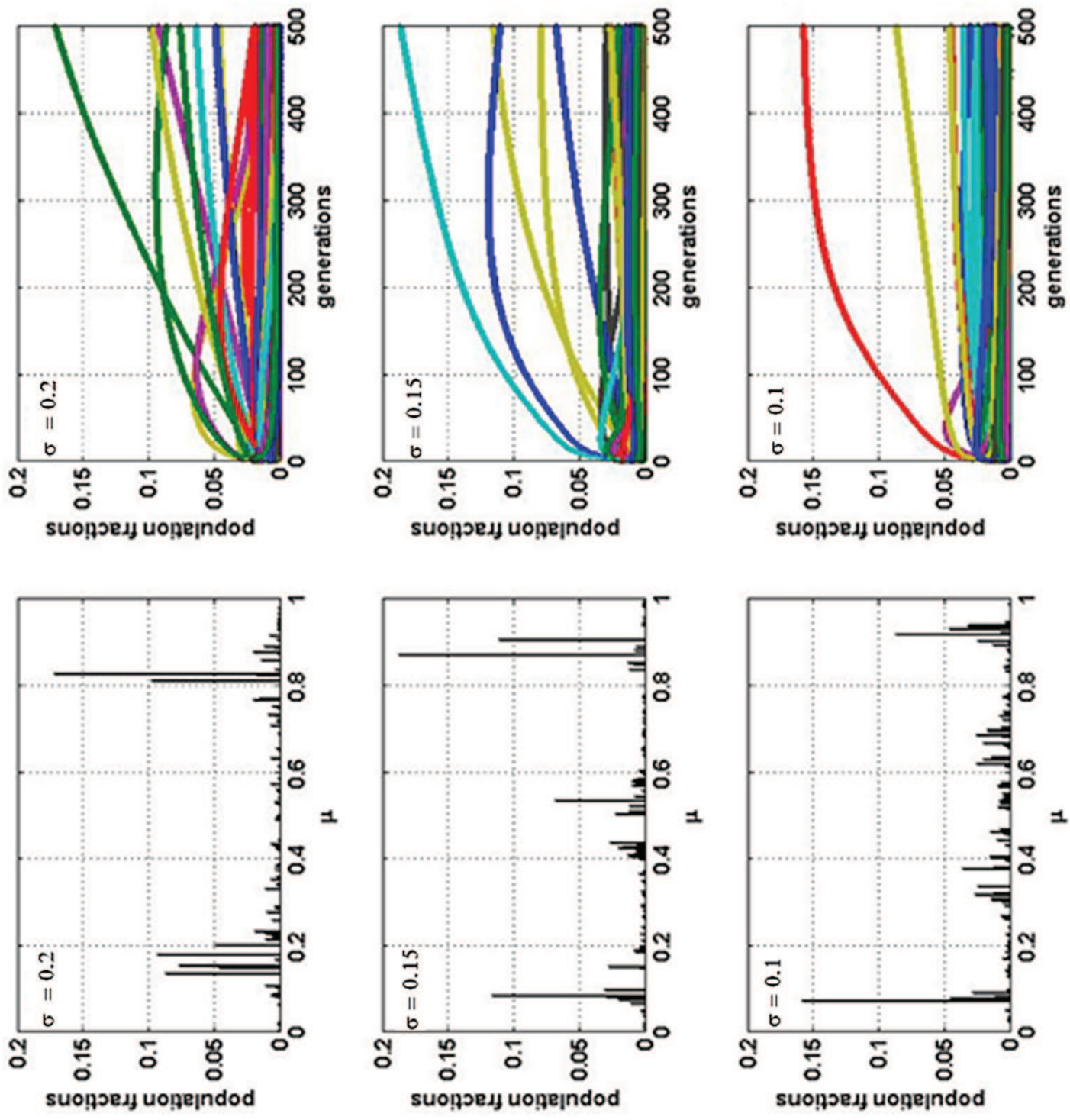

Figure 1

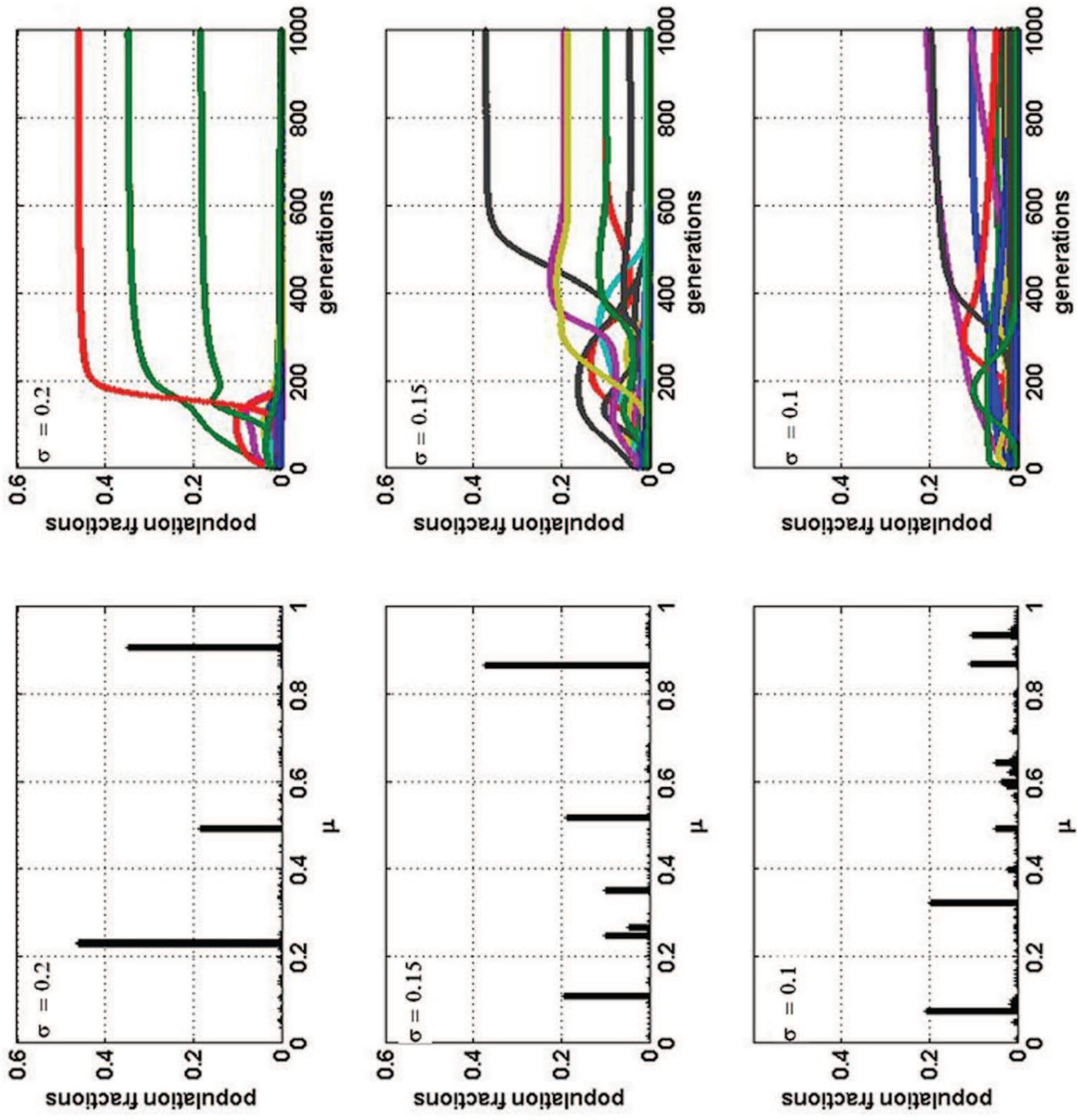

Figure 2

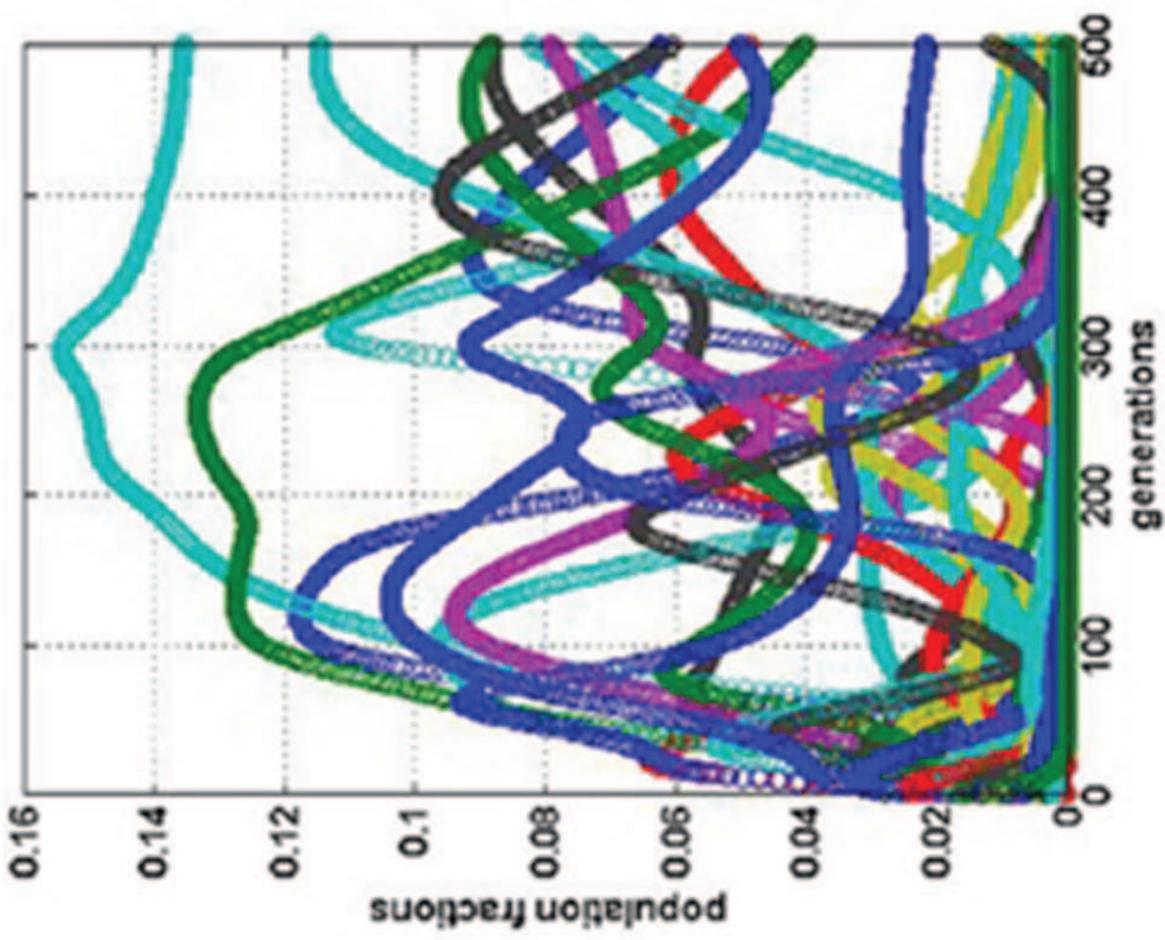
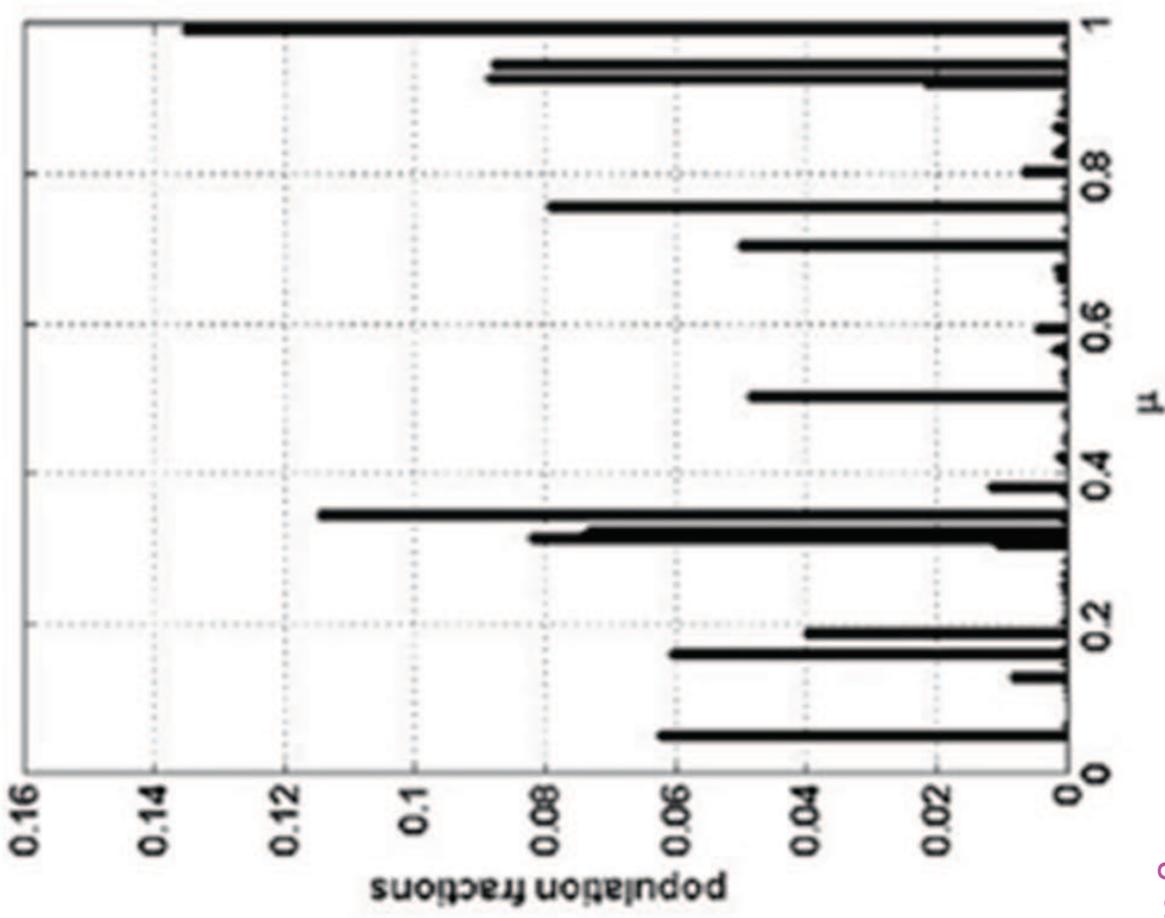

Figure 3

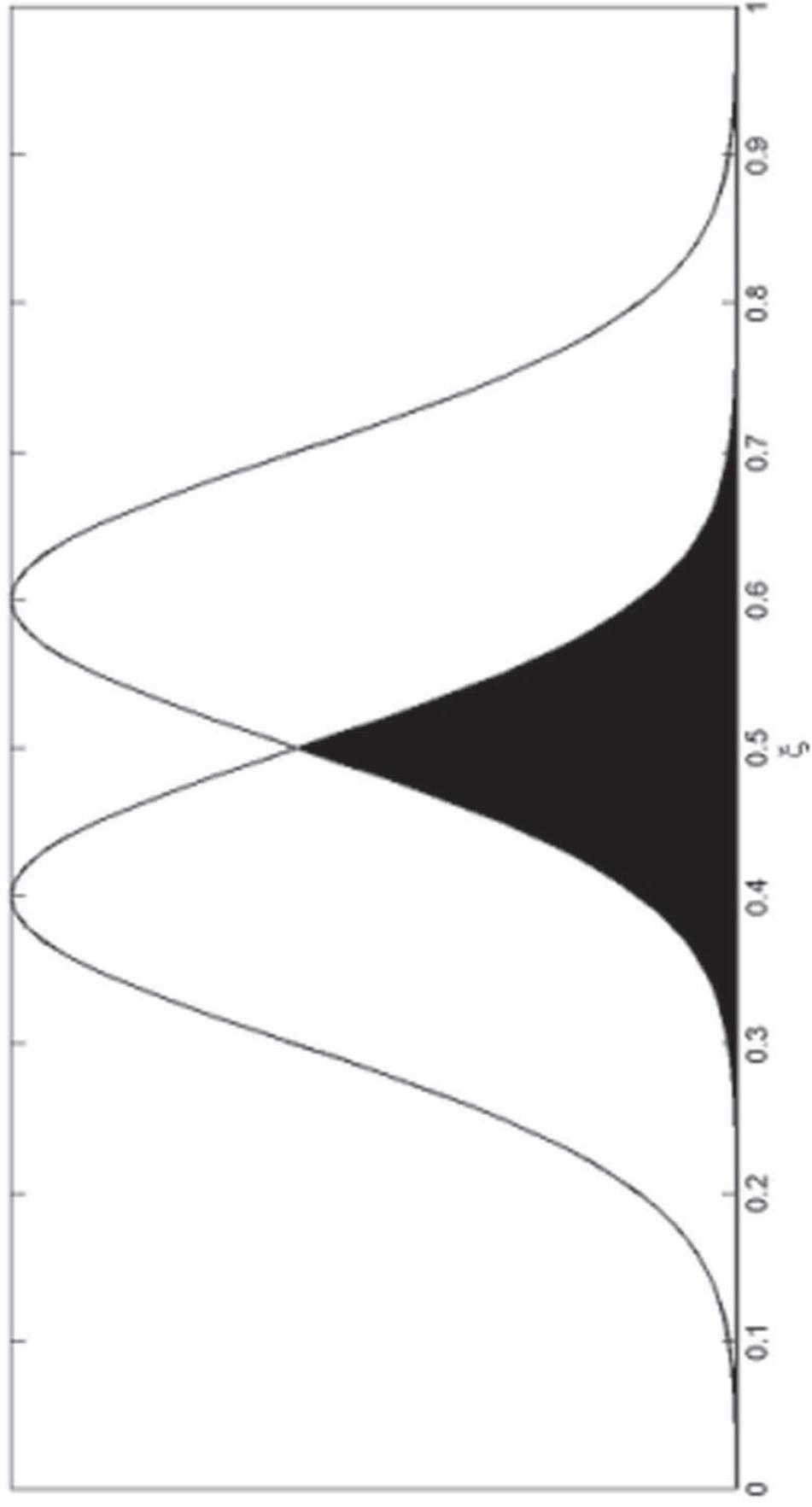

Figure 4

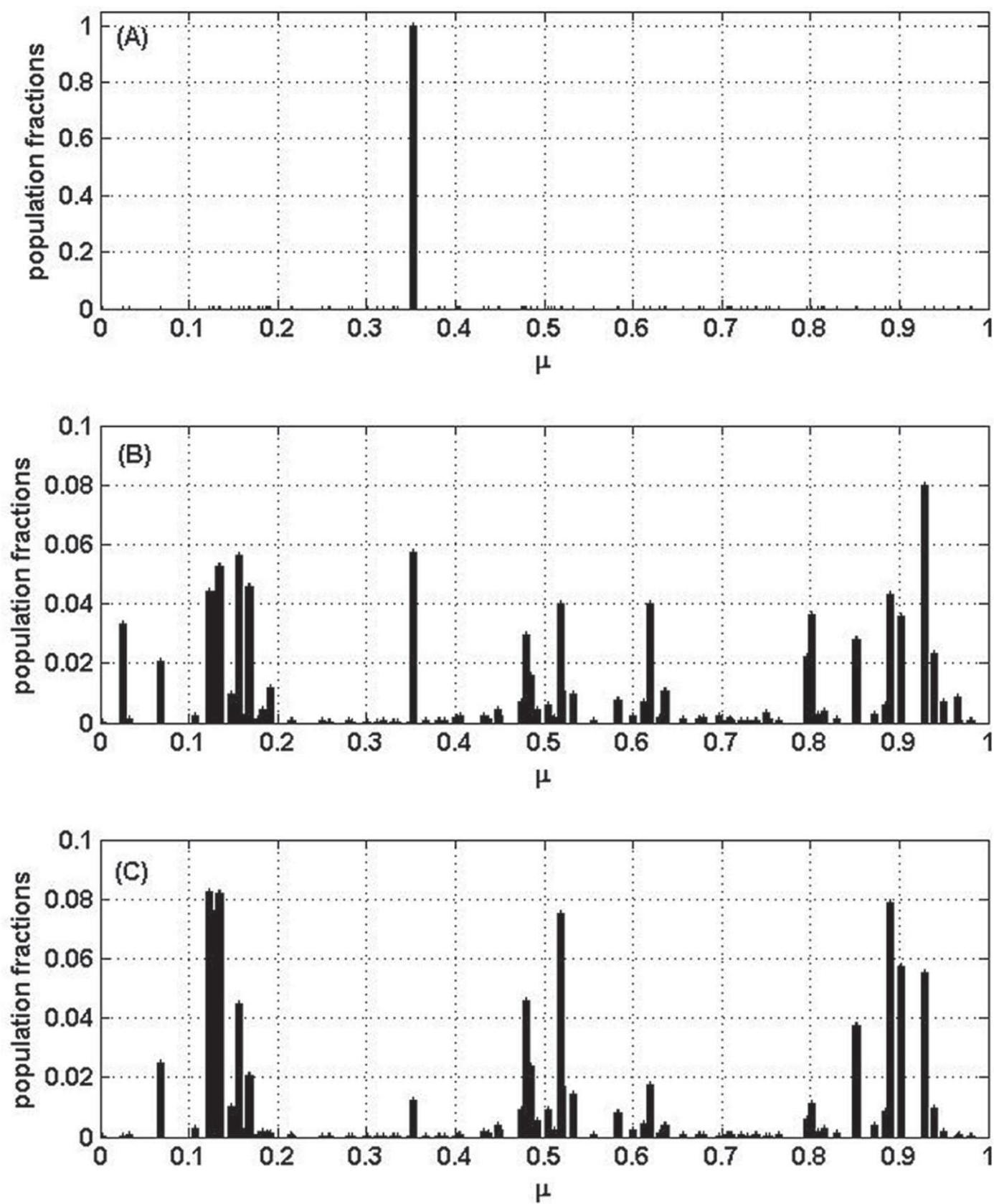

Figure 5